# Superconductivity in HEA-type compounds

Yoshikazu Mizuguchi*, Aichi Yamashita

*Department of Physics, Tokyo Metropolitan University, 1-1, Minami-osawa, Hachioji 192-0397, Japan*

**Abstract**

Since the discovery of superconductivity in a high-entropy alloy (HEA) Ti-Zr-Nb-Hf-Ta in 2014, the community of superconductor science has explored new HEA superconductors to find the merit of the HEA states on superconducting properties. Since 2018, we have developed "HEA-type" compounds as superconductors or thermoelectric materials. As well known, compounds like intermetallic compounds or layered compounds are composed of multi crystallographic sites. In a HEA-type compounds, one or more sites are alloyed and total mixing entropy satisfies with the criterion of HEA. Herein, we summarize the synthesis methods, the crystal structural variation and superconducting properties of the HEA-type compounds, which include NaCl-type metal tellurides, $CuAl_2$-type transition metal zirconides, high-$T_c$ cuprates, and $BiS_2$-based layered superconductors. The effects of the introduction of a HEA site in various kinds of complicated compounds are discussed from the structural-dimensionality viewpoint.

## 1. Introduction

### 1-1. Superconducting materials

Superconductivity is a quantum phenomenon, which is characterized by zero-resistivity states in electrical resistivity and exclusion of magnetic flux from a superconductor [1,2]. Superconductivity has provided many exotic research topics not only in the field of science but also in application of superconductors. The zero-resistivity states can achieve large-scale electricity transport with ultra-low energy loss, very high magnetic fields, which has been used in various devices like a magnetic resonance imaging (MRI) and a superconducting Maglev train. Although superconductor devises look perfect, the use of superconductors are regulated by temperature in reality because superconducting states are observed only at temperatures below a superconducting transition temperature ($T_c$), which is a parameter unique to the superconductor. To use superconducting devices, the system must be cooled down to low temperatures lower than the $T_c$ of its superconducting components. Therefore, discovery of high-$T_c$ superconductors has been desired.

In 1986, superconductivity with a high $T_c$ in a Cu oxide $(La,Ba)_2CuO_4$ was discovered [3,4]. Soon after the discovery, $T_c$ of the Cu-oxide superconductor (cuprate) family reached 90 K for $RE$$Ba_2Cu_3O_{7-d}$ ($RE$: rare earth) [5], which is higher than liquid nitrogen temperature, and finally reached 135 K in a Hg-Ba-Ca-Gu-O system [6]. After the discovery of the cuprate family, many layered compounds have been searched for high-$T_c$ superconductivity. In 2001, superconductivity in $MgB_2$ with a $T_c$ of 39 K was reported [7]. Furthermore, in 2008, FeAs-based layered superconductors $RE$$FeAsO_{1-x}F_x$ with a $T_c$ exceeding 50 K were discovered [8,9]. Particularly, in the cuprates and FeAs-based families, unconventional (non-phonon-mediated) mechanisms of superconductivity has been proposed to explain their high $T_c$ [10].

A surprising discovery of superconductivity at very high $T_c$ of 203 K in $H_3S$ was reported in 2015 [11]. The phenomenon could be achieved at extremely high pressures above 150 GPa, the high $T_c$ and possible conventional (phonon-mediated) mechanisms have recently been attracting many researchers in the field of condensed matter physics. Furthermore, higher $T_c$s have been reported in related hydrides, $LaH_{10}$ ($T_c$ >250 K at ~ 200 GPa) [12,13] and a carbonaceous sulfur hydride system ($T_c$ = 287.7 K at 267 GPa) [14]. To use high-$T_c$ superconducting states in hydrides, discovery of new superconducting hydrides which become superconductive at ambient pressure. The possible strategy to realize that is the utilization of chemical pressure effects, which are applied via chemical substitutions and sometimes work like external pressure effects. Therefore, further investigations on chemical pressure effects in novel superconductors are needed, and high-entropy alloying of compounds would be one of the routs to chemically modify the crystal structure of superconductors.

### 1-2. Superconductivity in high-entropy alloys

Although layered compounds had been the central topic in searching high-$T_c$ and/or unconventional superconductors over the last three decades, recent works on new superconductors have focused on various kinds of materials, which includes complicated compounds and pure metals as well. Among them, high-entropy-alloy (HEA) superconductors have been a developing field of study [15].

HEA is an alloy possessing high configurational mixing entropy ($\Delta S_{mix}$), which is achieved by making the alloy with more than five constituent elements with an occupancy ranging 5 to 35 at% for each element [16,17]. Typically, $\Delta S_{mix}$ of HEA is calculated by $\Delta S_{mix} = -R \Sigma_i c_i \ln c_i$, where $c_i$ and $R$ are the compositional ratio and the gas constant [17], and reaches 1.5$R$. Due to high $\Delta S_{mix}$, HEAs exhibit stability or high performance in high temperature and/or extreme conditions [17]. Therefore, HEAs have been extensively studied in the fields of materials science and engineering.

In 2014, Koželj et al. reported superconductivity with $T_c$ = 7.3 K in a HEA $Ta_{0.34}Nb_{0.33}Hf_{0.08}Zr_{0.14}Ti_{0.11}$ [18]. The HEA superconductor has a bcc structure with a space group of *Im*-3*m*. In Fig. 1, we compare the crystal structures of (a) a pure Nb metal ($T_c$ = 9.2 K), (b) a NbTi alloy ($T_c$ ~ 10 K), which is the mostly-used practical superconductor, and (c) HEA $Ta_{0.34}Nb_{0.33}Hf_{0.08}Zr_{0.14}Ti_{0.11}$. All those materials show superconductivity and the crystal structure type is the same. The difference between them is the mixing entropy $\Delta S_{mix}$ and $T_c$. Although the $T_c$ of HEA is lower than that of the other two, it was surprising for researchers that such a disordered alloy exhibits superconductivity with bulk nature. After the discovery of Ref. 18, various HEA superconductors have been developed; material information [18–26] is listed in Table I.

As shown in Fig. 2, a superconducting transition was observed in $Ta_{0.34}Nb_{0.33}Hf_{0.08}Zr_{0.14}Ti_{0.11}$ [18]. The temperature dependence of electrical resistivity (Fig. 2(a)) shows metallic behavior but exhibits a relatively small residual resistivity ratio (*RRR*) at low temperatures. This would be due to the presence of disorder, and a similar trend has been observed in various HEA-type superconductors. Bulk superconductivity could be confirmed through specific heat measurement as shown in Fig. 2(b). From specific heat data, it has been found that most HEA superconductors exhibit conventional (phonon-mediated) pairing states.

$T_c$ of HEA superconductors show correlation with valence electron count (VEC) [15,21,27]. The type-A HEAs with lower VEC exhibit a dependence of $T_c$ on VEC similar to that for crystalline metals, alloys, and amorphous materials, and the $T_c$ of HEAs are intermediate between crystalline materials and amorphous. In contrast, the $T_c$ of the type-C HEAs with middle VEC shows opposite behavior to that for other forms. For the type-B HEAs, the trend of $T_c$ on VEC seems resembling that for other forms, but their $T_c$s are clearly lower than that for crystalline and amorphous materials. As mentioned in Ref. 15, there would be a clear effect of crystallinity on $T_c$ at the same VEC range, but the origin of the different trends between type-A, type-B, and type-C regimes have not been clarified from physical viewpoint.

Another notable character of HEA is a robustness of superconductivity to extremely high pressure. As reported in Ref. 20, the $T_c$ of Ta-Nb-Hf-Zr-Ti slightly increases by external pressure effect and the 10 K-class $T_c$ maintains under extreme pressures like 200 GPa. However, the robustness of superconductivity to extremely high pressure was reported for simpler NbTi with a clear increase in $T_c$ to 19.1 K at 261 GPa [28]. Therefore, the issue if HEA can improve the stability of superconductor under extremely high pressures has not been clarified.

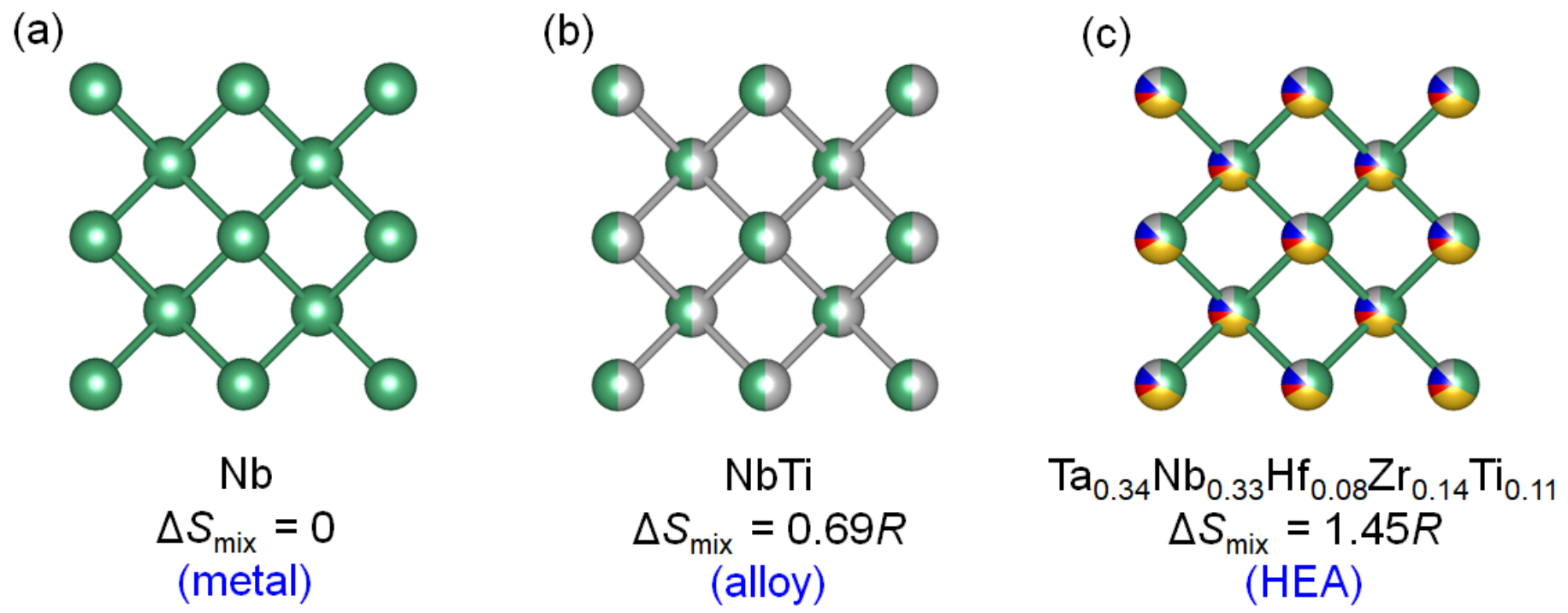

**Fig. 1. Schematic image of crystal structure of (a) Nb (pure metal), (b) NbTi (alloy), and (c) $Ta_{0.34}Nb_{0.33}Hf_{0.08}Zr_{0.14}Ti_{0.11}$ (HEA).**

**Table I. List of HEA superconductors and HEA-type superconducting compounds; composition, mixing entropy, $T_c$, and structural type are summarized.**

| Composition | $\Delta S_{mix}/R$ | $T_c$ (K) | Structure | Ref. |
|---|---|---|---|---|
| $Ta_{0.34}Nb_{0.33}Hf_{0.08}Zr_{0.14}Ti_{0.11}$ | 1.45 | 7.3 | bcc | [18][19] |
| $(TaNb)_{0.67}(HfZrTi)_{0.33}$ | 1.46 | 7.7 | bcc | [20] |
| $(TaNb)_{0.7}(ZrHfTi)_{0.3}$ | 1.43 | 8.0 | bcc | [21] |
| $(TaNb)_{0.7}(ZrHfTi)_{0.33}$ | 1.24 | 7.8 | bcc | |
| $(TaNb)_{0.7}(ZrHfTi)_{0.4}$ | 1.31 | 7.6 | bcc | |
| $(TaNb)_{0.7}(ZrHfTi)_{0.5}$ | 1.39 | 6.5 | bcc | |
| $(TaNb)_{0.7}(ZrHfTi)_{0.84}$ | 1.60 | 4.5 | bcc | |
| $(TaNb)_{0.67}(Hf)_{0.33}$ | 1.10 | 7.3 | bcc | |
| $(TaNb)_{0.67}(HfZr)_{0.33}$ | 1.33 | 6.6 | bcc | |
| $Nb_{0.67}(HfZrTi)_{0.33}$ | 1.00 | 9.2 | bcc | |
| $(NbV)_{0.67}(HfZrTi)_{0.33}$ | 1.46 | 7.2 | bcc | |
| $(TaV)_{0.67}(HfZrTi)_{0.33}$ | 1.46 | 4.0 | bcc | |
| $(TaNb)_{0.67}(HfZrTi)_{0.33}$ | 1.46 | 7.3 | bcc | |
| $(TaNbV)_{0.67}(HfZrTi)_{0.33}$ | 1.73 | 4.3 | bcc | |
| $Hf_{0.21}Nb_{0.25}Ti_{0.15}V_{0.15}Zr_{0.24}$ | 1.59 | 5.3 | bcc | [22] |
| $Ta_{0.35}Nb_{0.35}Zr_{0.15}Ti_{0.15}$ | 1.30 | 8.0 | bcc | [23] |
| $(ZrNb)_{0.2}(MoReRu)_{0.8}$ | 1.52 | 4.2 | bcc | |
| $(ZrNb)_{0.1}(MoReRu)_{0.9}$ | 1.38 | 5.3 | bcc | |
| $(HfTaWIr)_{0.6}Re_{0.4}$ | 1.50 | 1.9 | bcc+hcp | |
| $(HfTaWIr)_{0.5}Re_{0.5}$ | 1.39 | 2.7 | bcc+hcp | |
| $(HfTaWIr)_{0.4}Re_{0.6}$ | 1.23 | 4.0 | bcc | |
| $(HfTaWIr)_{0.3}Re_{0.7}$ | 1.03 | 4.5 | bcc | [24] |
| $(HfTaWIr)_{0.2}Re_{0.8}$ | 0.78 | 5.7 | bcc | |
| $(HfTaWPt)_{0.5}Re_{0.5}$ | 1.39 | 2.2 | bcc+hcp | |
| $(HfTaWPt)_{0.4}Re_{0.6}$ | 1.23 | 4.4 | bcc | |
| $(HfTaWPt)_{0.3}Re_{0.7}$ | 1.03 | 5.7 | bcc | |
| $(HfTaWPt)_{0.25}Re_{0.75}$ | 0.91 | 6.1 | bcc | |
| $Nb_{26.1}Ta_{25.1}Ti_{23.4}Zr_{0.254}$ | 1.39 | 8.3 | bcc | |
| $Nb_{0.198}Ta_{0.189}Ti_{0.208}Zr_{0.187}Hf_{0.218}$ | 1.61 | 7.1 | bcc | |
| $Nb_{0.163}Ta_{0.157}Ti_{0.169}Zr_{0.171}Hf_{0.175}V_{0.165}$ | 1.79 | 5.1 | bcc | |
| $Nb_{0.2}Ta_{0.2}Ti_{0.2}Zr_{0.2}Fe_{0.2}$ | 1.61 | 6.9 | bcc | [25] |
| $Nb_{0.2}Ta_{0.2}Ti_{0.2}Zr_{0.2}Ge_{0.2}$ | 1.61 | 8.4 | bcc | |
| $Nb_{0.2}Ta_{0.2}Ti_{0.2}Zr_{0.2}Si_{0.2}V_{0.2}$ | 1.79 | 4.3 | bcc | |
| $Nb_{0.2}Ta_{0.2}Ti_{0.2}Zr_{0.2}Si_{0.2}Ge_{0.2}$ | 1.79 | 7.4 | bcc | |
| $(TaNb)_{0.31}(TiUHf)_{0.69}$ | 1.59 | 3.2 | bcc | [26] |

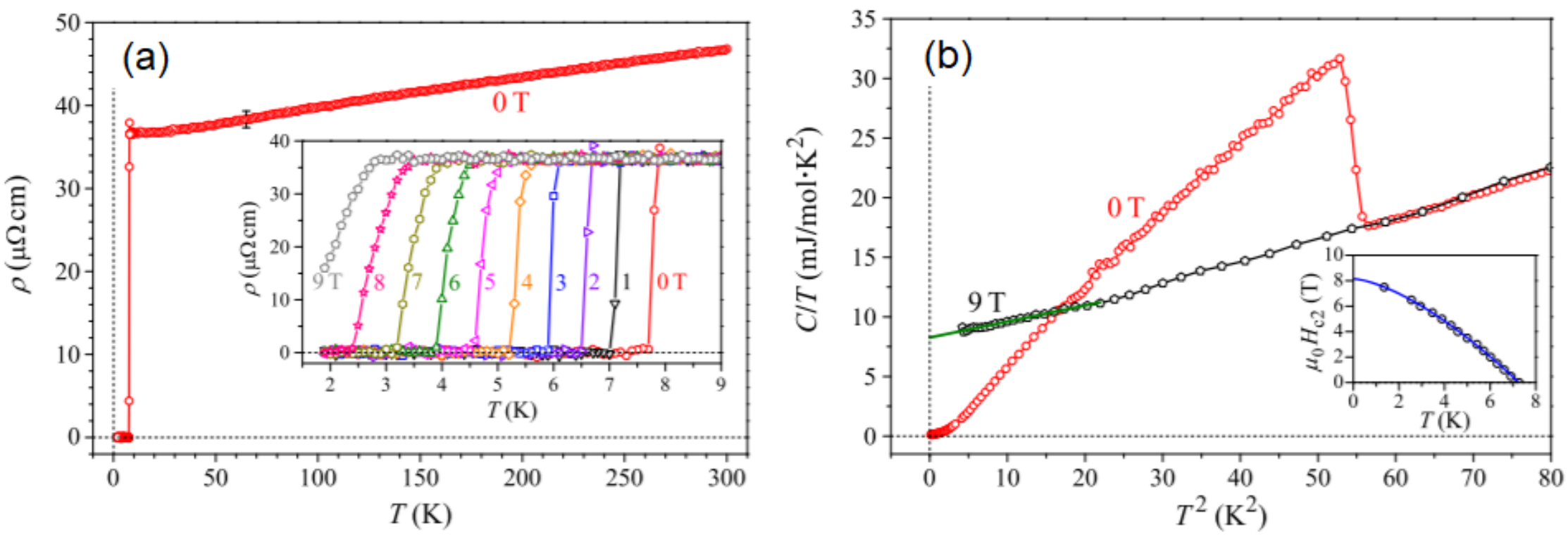

**Fig. 2. (a) Temperature dependence of electrical resistivity ($\rho$) of $Ta_{0.34}Nb_{0.33}Hf_{0.08}Zr_{0.14}Ti_{0.11}$. (b) $T^2$ dependence of $C/T$, where $C$ is specific heat of $Ta_{0.34}Nb_{0.33}Hf_{0.08}Zr_{0.14}Ti_{0.11}$. The figures were reproduced under permission by the authors of Ref. 18 (DOI: 10.1103/PhysRevLett.113.107001) and APS. Copyright 2014 by American Physical Society.**

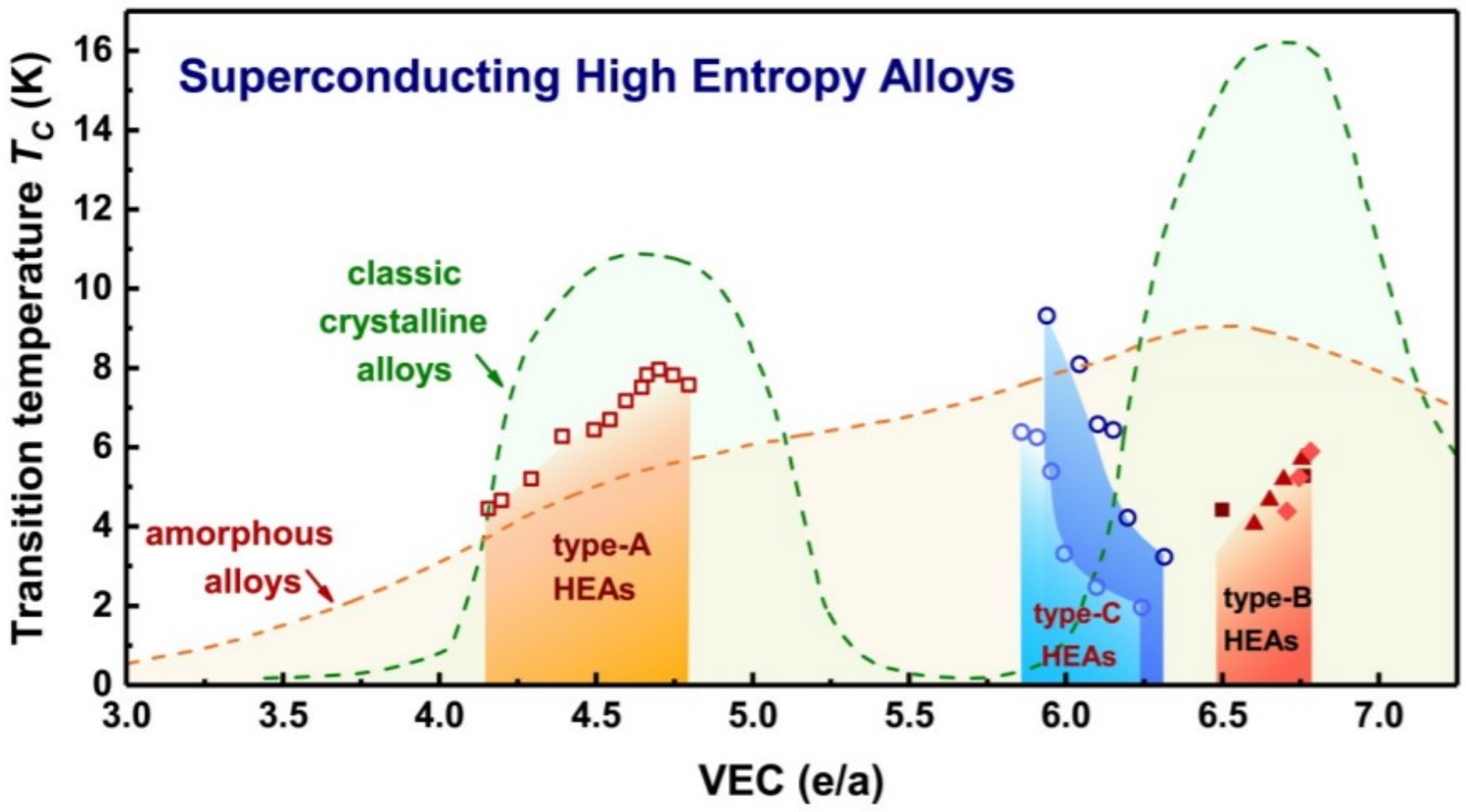

**Fig. 3. Valence electron count (VEC) dependence of $T_c$ for classic crystalline alloys, amorphous alloys, and HEAs. The figure was reproduced under permission by the authors of Ref. 15 (10.1103/PhysRevMaterials.3.090301) and APS. Copyright 2019 by American Physical Society.**

## 1-3. Concept of HEA-type compounds

As described in subsection 1-2, superconductivity in HEAs has been discovered and been regarded as a new research field of superconductivity. However, the merit of HEA states for superconductors has not been fully understood. Therefore, development of new types of HEA superconductors is needed. The hint to expand the material variation of HEA superconductors was proposed in Ref. 29, in which a HEA superconductor with a CsCl-type structure was reported. Since the CsCl structure contains two independent crystallographic sites, we have flexibility of elemental solution at the two sites. When calculating total $\Delta S_{mix}$ of $(ScZrNbTa)_{0.65}(RhPd)_{0.35}$ by taking the sum of $\Delta S_{mix}$s for site-1 and site-2, it appears to reach very high $\Delta S_{mix}$ of 1.79*R*. A similar site separation has been observed in $(Nb_{0.11}Re_{0.56})(HfZrTi)_{0.33}$ [30]. Motivated by those studies on HEAs with site separation, we have tried to synthesize various "*HEA-type compounds*", which contain NaCl-type metal chalcogenides [31-33], $CuAl_2$-type tetragonal $TrZr_2$ (*Tr*: Fe, Co, Ni, Cu, Rh, Ir) [34,35], high-$T_c$ RE123 cuprates [36], and $BiS_2$-based layered superconductors [37,38]. The concept of HEA-type compounds it that we achieve a high $\Delta S_{mix}$ by site-selective alloying. As shown in Fig. 4, HEA-type compounds have a HEA-type site, in which five or more elements are solving and a normal site, which is not in the HEA state. The list of superconducting HEA-type compounds is shown in Table II. By studying HEA effects to crystal structure and physical properties in various crystal structures, we could identify the merit of HEA states in those compounds. In section 2, we review the material synthesis, crystal structure, and physical properties of newly synthesized HEA-type compounds.

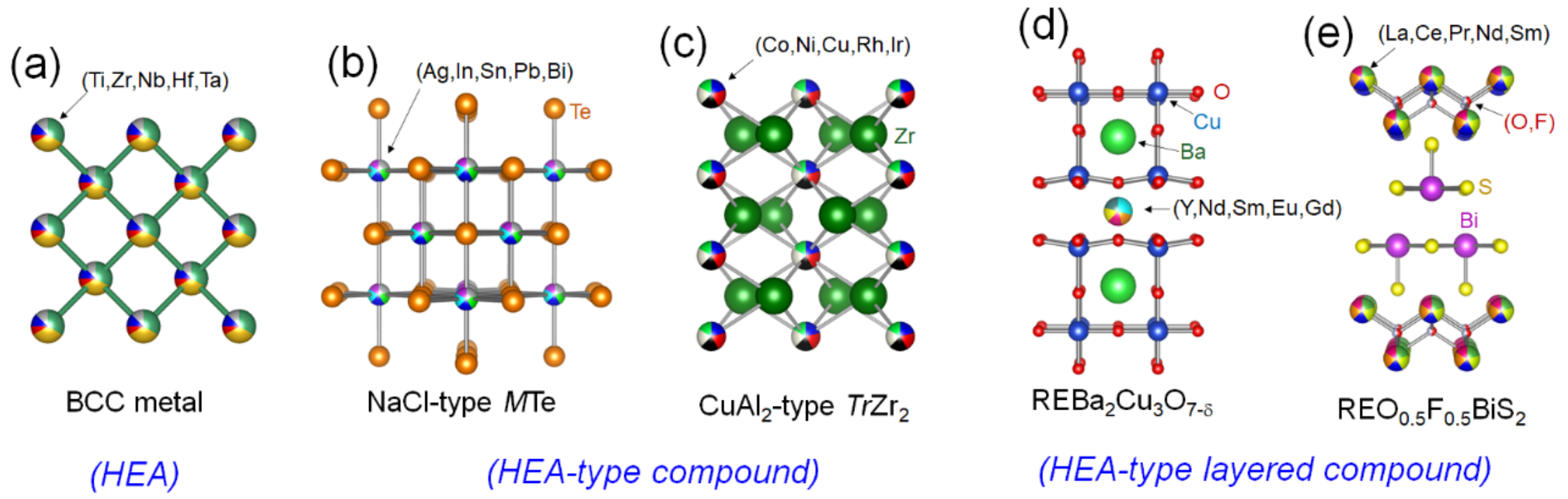

**Fig. 4. Crystal structure images of conventional HEA (a) and HEA-type compounds (b–e).**

**Table II. List of HEA-type superconducting compounds; composition, mixing entropy (site-1, site-2, total), $T_c$, and structural type are summarized.**

| Composition | $\Delta S_{mix}/R$ (site1) | $\Delta S_{mix}/R$ (site2) | $\Delta S_{mix}/R$ (Total) | $T_c$ (K) | Structure | Ref. |
|---|---|---|---|---|---|---|
| $(ScZrNbTa)_{0.65}(RhPd)_{0.35}$ | 1.18 | 0.61 | 1.79 | 9.3 | CsCl $Pm3m$ | [29] |
| $(ScZrNb)_{0.63}(RhPd)_{0.37}$ | 1.16 | 0.62 | 1.79 | 7.5 | | |
| $(ScZrNb)_{0.62}(RhPd)_{0.38}$ | 1.16 | 0.63 | 1.79 | 6.4 | | |
| $(ScZrNb)_{0.60}(RhPd)_{0.40}$ | 1.14 | 0.64 | 1.78 | 3.9 | | |
| $(Nb_{0.11}Re_{0.56})(HfZrTi)_{0.33}$ | 0.73 | 0.57 | 1.30 | 4.4 | hcp | [30] |
| $(Ag_{0.2}In_{0.2}Sn_{0.2}Pb_{0.2}Bi_{0.2})Te$ | 1.61 | 0 | 1.61 | 2.8 | NaCl $Fm$-$3m$ | [31] |
| $(Ag_{0.20}Cd_{0.20}Sn_{0.20}Sb_{0.15}Pb_{0.20})Te_{1.05}$ | 1.60 | 0 | 1.60 | 1.2 | NaCl $Fm$-$3m$ | [32] |
| $(Ag_{0.24}In_{0.22}Sn_{0.18}Sb_{0.14}Pb_{0.19})Te_{1.03}$ | 1.59 | 0 | 1.59 | 1.4 | | |
| $(Ag_{0.22}Cd_{0.22}In_{0.23}Sn_{0.17}Sb_{0.14})Te_{1.02}$ | 1.59 | 0 | 1.59 | 0.7 | | |
| $(Ag_{0.19}Cd_{0.19}Sn_{0.20}Pb_{0.18}Bi_{0.21})Te_{1.03}$ | 1.61 | 0 | 1.61 | 1.0 | | |
| $(Ag_{0.21}Cd_{0.19}In_{0.25}Pb_{0.16}Bi_{0.18})Te_{1.00}$ | 1.60 | 0 | 1.60 | 1.0 | | |
| $(Ag_{0.21}Cd_{0.21}In_{0.24}Sn_{0.19}Bi_{0.19})Te_{0.97}$ | 1.61 | 0 | 1.61 | 1.0 | | |
| $(Ag_{0.24}In_{0.22}Pb_{0.27}Bi_{0.26})Te_{1.02}$ | 1.37 | 0.00 | 1.37 | 2.7 | NaCl $Fm$-$3m$ | [33] |
| $(Ag_{0.29}In_{0.26}Pb_{0.22}Bi_{0.24})(Te_{0.78}Se_{0.20})$ | 1.38 | 0.51 | 1.89 | 2.5 | | |
| $(Ag_{0.34}In_{0.15}Pb_{0.24}Bi_{0.29})(Te_{0.65}Se_{0.34})$ | 1.35 | 0.65 | 2.00 | 2.0 | | |
| $(Co_{0.2}Ni_{0.1}Cu_{0.1}Rh_{0.3}Ir_{0.3})Zr_2$ | 1.47 | 0 | 1.47 | 7.8 | $CuAl_2$ $I4/mcm$ | [34] |
| $(Fe_{0.093}Co_{0.194}Ni_{0.113}Rh_{0.271}Ir_{0.329})Zr_2$ | 1.50 | 0 | 1.50 | 7.8 | $CuAl_2$ $I4/mcm$ | [35] |
| $(Fe_{0.108}Co_{0.297}Ni_{0.202}Rh_{0.073}Ir_{0.320})Zr_2$ | 1.48 | 0 | 1.48 | 6.7 | | |
| $(Fe_{0.190}Co_{0.190}Ni_{0.200}Rh_{0.212}Ir_{0.208})Zr_2$ | 1.61 | 0 | 1.61 | 5.4 | | |
| $(Fe_{0.293}Co_{0.190}Ni_{0.300}Rh_{0.093}Ir_{0.124})Zr_2$ | 1.52 | 0 | 1.52 | 4.8 | | |
| $(Y_{0.28}Nd_{0.16}Sm_{0.18}Eu_{0.18}Gd_{0.20})Ba_2Cu_3O_{7-d}$ | 1.59 | 0 | 1.59 | 93.0 | Layered *Pmmm* | [36] |
| $(Y_{0.18}La_{0.24}Nd_{0.14}Sm_{0.14}Eu_{0.15}Gd_{0.15})Ba_2Cu_3O_{7-d}$ | 1.77 | 0 | 1.77 | 93.0 | | |
| $(La_{0.2}Ce_{0.2}Pr_{0.2}Nd_{0.2}Sm_{0.2})O_{0.5}F_{0.5}BiS_2$ | 1.61 | 0.69 | 2.30 | 4.3 | Layered *P*4/*nmm* | [37][38] |
| $(La_{0.3}Ce_{0.3}Pr_{0.2}Nd_{0.1}Sm_{0.1})O_{0.5}F_{0.5}BiS_2$ | 1.50 | 0.69 | 2.20 | 3.4 | Layered *P*4/*nmm* | [37] |
| $(La_{0.1}Ce_{0.1}Pr_{0.3}Nd_{0.3}Sm_{0.2})O_{0.5}F_{0.5}BiS_2$ | 1.50 | 0.69 | 2.20 | 4.7 | | |
| $(La_{0.1}Ce_{0.1}Pr_{0.2}Nd_{0.3}Sm_{0.3})O_{0.5}F_{0.5}BiS_2$ | 1.50 | 0.69 | 2.20 | 4.9 | | |
| $(La_{0.28}Ce_{0.32}Pr_{0.21}Nd_{0.09}Sm_{0.10})BiS_2$ | 1.50 | 0 | 1.50 | 3.4 | Layered *P*4/*nmm* | [39] |
| $(La_{0.10}Ce_{0.29}Pr_{0.33}Nd_{0.19}Sm_{0.09})BiS_2$ | 1.49 | 0 | 1.49 | 4.3 | | |
| $(La_{0.23}Ce_{0.21}Pr_{0.19}Nd_{0.19}Sm_{0.17})BiS_2$ | 1.60 | 0 | 1.60 | 3.3 | | |
| $(La_{0.09}Ce_{0.29}Pr_{0.12}Nd_{0.21}Sm_{0.29})BiS_2$ | 1.52 | 0 | 1.52 | 4.6 | | |

## 2. NaCl-type metal chalcogenides *MCh*

### 2-1. Metal tellurides *M*Te

The NaCl-type metal telluride family is one of the hot systems because it includes thermoelectric PbTe [40] and a topological crystalline insulator SnTe [41]. For superconducting tellurides, high-pressure synthesis was used to stabilize the NaCl-type structure [42–44]. For example, the low-pressure phase of InTe has a TlSe-type structure, but the high-pressure phase of InTe has a NaCl-type structure. The high-pressure phase can be obtained by high-pressure synthesis [43,44]. Motivated by these facts, we tried to synthesize HEA-type tellurides *M*Te where the *M* site is in the HEA state (see Fig. 4(b) for crystal structure) by high-pressure synthesis.

Figure 5(a) shows the temperature dependence of electrical resistivity for $AgInSnPbBiTe_5$, in which the *M* site is evenly occupied by Ag, In, Sn, Pb, and Bi (five metals) [31]. Very small *RRR* was observed, which is a similar trend to that in HEA superconductors [18]. In addition, four different *M*Te (*M*: Ag, In, Cd, Sn, Sb, Pb, Bi) superconductors with a HEA-type site has been obtained [32]. Interestingly, there is a correlation between the lattice constant and $T_c$ in HEA-type *M*Te. In Figure 5(b), the data for typical *M*Te superconductors are plotted. It is found that the trend that $T_c$ increases with increasing lattice constant is common among the plotted *M*Te. The $T_c$s of HEA-type are, however, lower than those of low-entropy tellurides, such as InTe and (In,Sn)Te. Therefore, the introduction of the HEA-type *M* site is found to negatively work for $T_c$ in MTe. This negative effect would be due to the direct effect of strong disorder to the *M*-Te bonding states and hence electronic states.

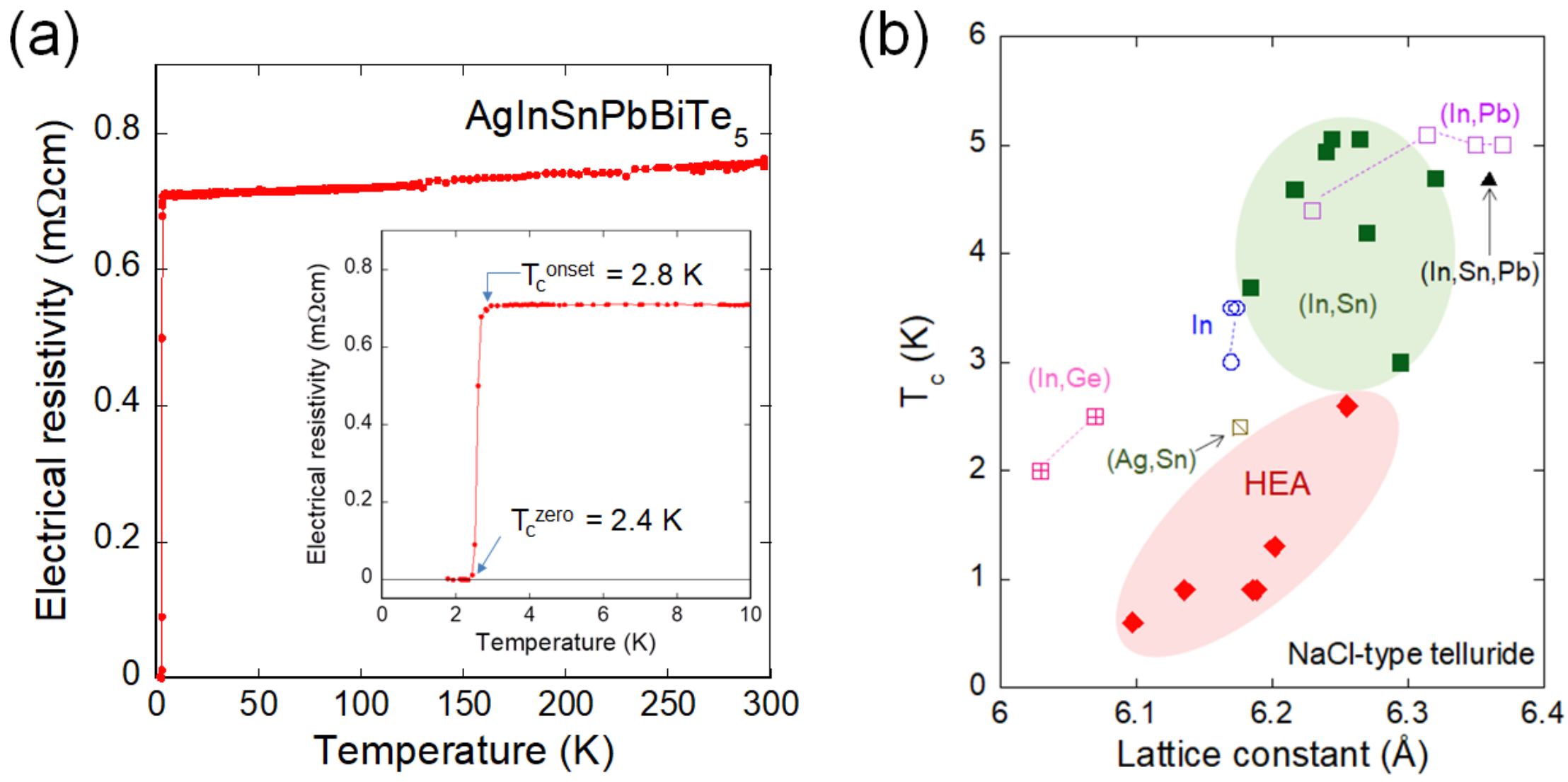

**Fig. 5. (a) Temperature dependence of the electrical resistivity of AgInSnPbBiTe$_5$. Original data has been published in Ref. 31. Copyright 2019 by the Physical Society of Japan. (b) Relationship between lattice constant and $T_c$ for metal tellurides including HEA-type tellurides. The elements written in the figure indicate compositions of the *M* site. Original data has been published in Ref. 32. Copyright 2020 by IOP.**

2-2. Hybrid high-entropy alloying in *MCh*

The Te site of *M*Te can be substituted by S and Se. The flexibility of both *M* and Te sites to element substitution enables us to design "*hybrid HEA*", in which both sites are alloyed [33]. Figure 6(a) shows the X-ray diffraction patterns for $(Ag_{0.25}In_{0.25}Pb_{0.25}Bi_{0.25})Te_{1-x}Se_x$. Notably, mixing many elements at two sites does not result in phase separation, and a single-phase sample was obtained for $x = 0.25$, while small impurity phases were detected for $x = 0.5$. For $x = 0.25$, as displayed in Fig. 6(b), the $\Delta S_{mix}$ for the *M* and *Ch* sites are 1.38*R* and 0.51*R*, and the total $\Delta S_{mix}$ reaches 1.89*R*. Furthermore, the total $\Delta S_{mix}$ for $x = 0.5$ reaches 2.00*R*. These $\Delta S_{mix}$ values are clearly higher than that for HEAs (Table I). Therefore, alloying at two or more sites (hybrid high-entropy alloying) results in very high total $\Delta S_{mix}$.

The superconducting properties for $(Ag_{0.25}In_{0.25}Pb_{0.25}Bi_{0.25})Te_{1-x}Se_x$ are shown in Fig. 6(c). A superconducting transition with $T_c > 2$ K was observed for $x = 0$ and 0.25. By focusing on these two phases, interesting trend was found. Although the $T_c$ for $x = 0.25$ is lower than that for $x = 0$, the suppression of $T_c$ for $x = 0.25$ under magnetic fields is clearly smaller than that for $x = 0$. From the estimation of upper critical field ($H_{c2}$), it was found that the $H_{c2}$ (0 K) for $x = 0.25$ is higher than that for $x = 0$. In addition, from the measurements of the magnetization-field loop, it was confirmed that the critical current density ($J_c$) at 1.8 K for $x = 0.25$ is larger than that for $x = 0$. These results suggest that an increase in $\Delta S_{mix}$ may be useful to improve $H_{c2}$ and/or $J_c$ characteristics of superconductors if the problem on the suppression of $T_c$ could be solved.

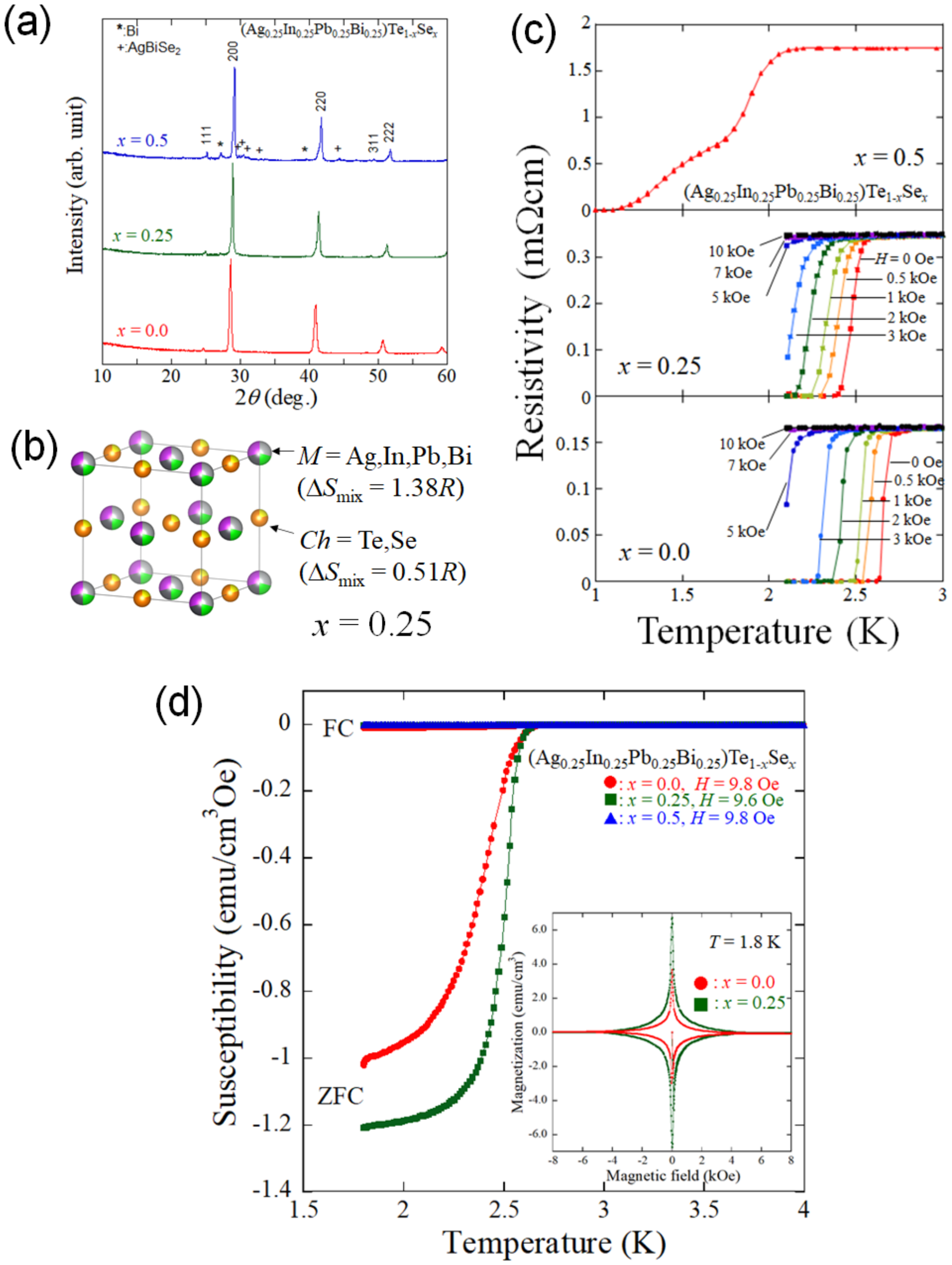

**Fig. 6. (a) X-ray diffraction patterns for hybrid-HEA-type $(Ag_{0.25}In_{0.25}Pb_{0.25}Bi_{0.25})Te_{1-x}Se_x$. (b) Schematic image of the crystal structure and mixing entropy for $x$ = 0.25 (c) Temperature dependences of electrical resistivity $(Ag_{0.25}In_{0.25}Pb_{0.25}Bi_{0.25})Te_{1-x}Se_x$ under magnetic fields. (d) Temperature dependence of magnetic susceptibility. The inset shows the magnetization-field curves. Original data has been published in Ref. 33. Copyright 2020 by the Royal Society of Chemistry.**

## 3. $CuAl_2$-type *Tr*$Zr_2$

As shown in the last section, high-entropy alloying of a compound, which possesses two or more crystallographic sites, is a route to develop new superconductors with a high mixing entropy. In addition, in HEA-type compounds, not only the site entropy but also the entropy of chemical bonding states should be higher than normal alloys or compounds. To discover new HEA-type superconductors, the use of material database is quite useful. SuperCon (NIMS database) [45] is a database of superconductors and contain information of composition, structural type, $T_c$, and reference of the material. To achieve the material design of HEA-type superconductors, we have to find a system in which compositional variation is rich within the same structural type, and superconducting transition has been observed. Herein, we introduce an example of material design for new HEA-type superconductors.

*Tr*$Zr_2$ ($T_r$: transition metal) with a tetragonal $CuAl_2$-type structure (Fig. 4(c)) is a superconducting system. For *Tr* = Fe, Co, Ni, Rh, Ir, superconductivity was reported. Among them, $CoZr_2$, $RhZr_2$, and $IrZr_2$ exhibits a higher $T_c$ of 5.5, 11.3, and 7.5 K, respectively [46]. Furthermore, the *Tr* site can be partially substituted by Sc, Cu, Ga, Pd, and Ta [34,45]. These facts suggest that the *Tr* site of *Tr*$Zr_2$ can be modified into a HEA-type site. We synthesized $Co_{0.2}Ni_{0.1}Cu_{0.1}Rh_{0.3}Ir_{0.3}Zr_2$ by arc melting and observed superconductivity with a $T_c$ of 8 K [34]. The *Tr* site of $Co_{0.2}Ni_{0.1}Cu_{0.1}Rh_{0.3}Ir_{0.3}Zr_2$ contains five transition metals, and the $\Delta S_{mix}$ for the *Tr* site is about 1.5*R*. In addition, (Fe,Co,Ni,Rh,Ir)$Zr_2$ superconductors were also synthesized and exhibited superconductivity [35]. Interestingly, the resulting $T_c$ in HEA-type phases was very close to the weighted-average $T_c$ of pure *Tr*$Zr_2$ systems. Although the origin of the unexpected behavior is still unclear, the effects of disordering by the HEA-type site to superconducting properties seem very limited. The difference in the HEA effects between *MCh* and *Tr*$Zr_2$ may be caused by the different structural complexity.

As shown in Fig. 7(a), the temperature dependence of resistivity exhibits a relatively large *RRR* as compared to other HEA-type superconductors. However, a large *RRR* of ~30 was reported for $CoZr_2$ single crystals [47], which is clearly larger than that for HEA-type *Tr*$Zr_2$. Therefore, the HEA effects (disordering effects) to transport properties would be a common trend. Figure 7(b) shows the temperature dependences of the specific heat for $Co_{0.2}Ni_{0.1}Cu_{0.1}Rh_{0.3}Ir_{0.3}Zr_2$ in the form of $C/T$. Although a clear jump of $C/T$ at the $T_c$ is seen, the transition is broader as compared to the case of $CoZr_2$. The trend of broad transition in specific heat at $T_c$ was not observed in a HEA superconductor but observed in HEA-type compounds. Because the superconducting transitions observed in resistivity and magnetization were sharp, the broad transition in the specific heat would suggest microscopic local phase separation in HEA-type superconducting compounds.

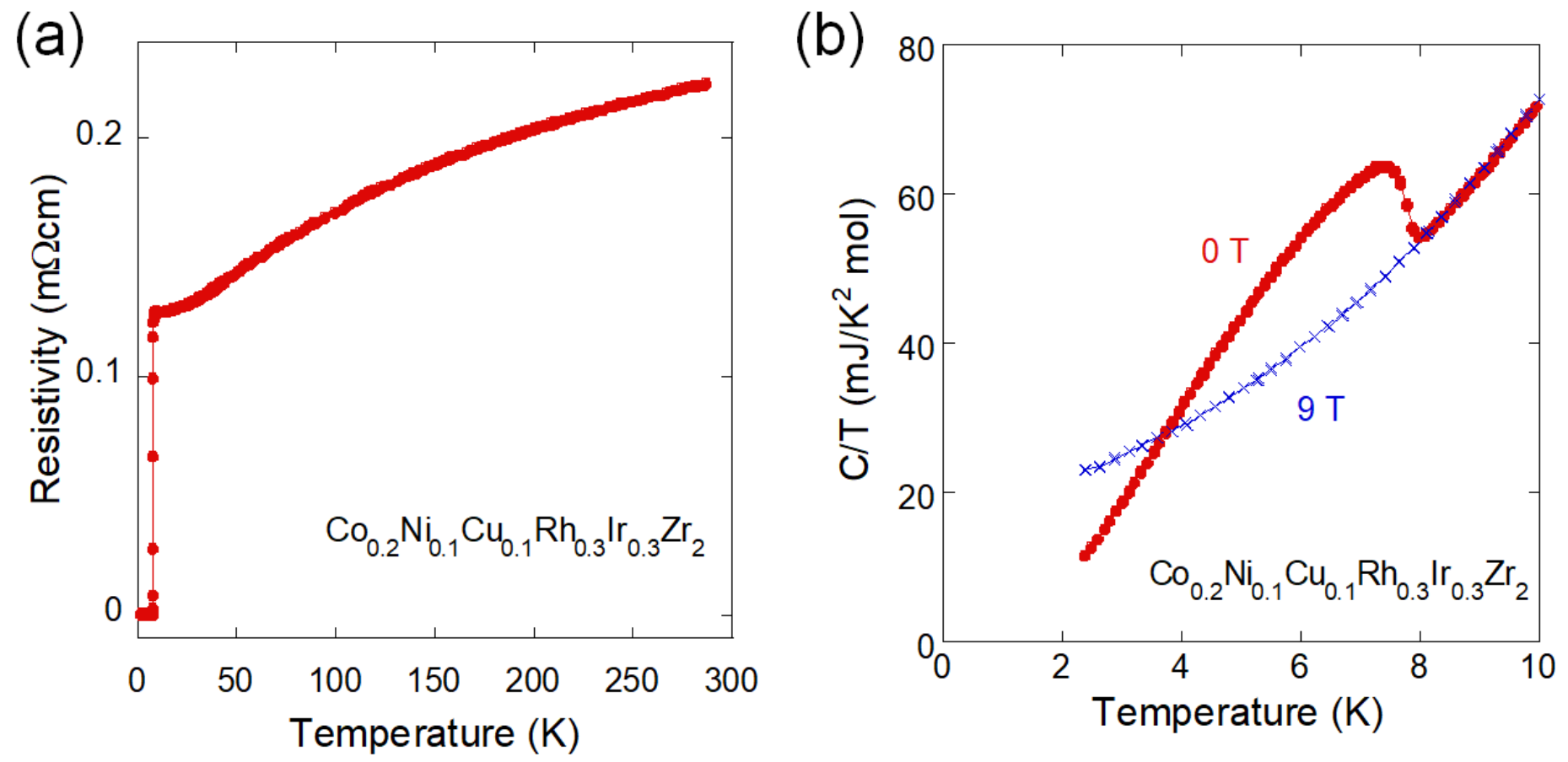

**Fig. 7. (a) Temperature dependence of resistivity for $Co_{0.2}Ni_{0.1}Cu_{0.1}Rh_{0.3}Ir_{0.3}Zr_2$. (b) Temperature dependence of specific heat ($C/T$) for for $Co_{0.2}Ni_{0.1}Cu_{0.1}Rh_{0.3}Ir_{0.3}Zr_2$ at 0 and 9 T. Original data has been published in Refs. 34 and 35. Copyright 2020 by Taylor & Francis.**

## 4. Cuprate (high-$T_c$) superconductors $RE$Ba$_2$Cu$_3$O$_{7-d}$

As mentioned in introduction, cuprates (Cu oxides) have been extensively studied in the fields of science and engineering because of its high $T_c$. Among them, $RE$Ba$_2$Cu$_3$O$_{7-d}$ ($RE$123) system [5] is one of practical materials for superconductivity application. In addition, a high $J_c$ was reported in $RE$123 samples with three elements at the $RE$ site [48]. Motivated by the fact, we synthesized polycrystalline samples of $RE$Ba$_2$Cu$_3$O$_{7-d}$ with different $\Delta S_{mix}$ for the $RE$ site [36] by standard solid-state reaction in air. In the study, two-step annealing was performed to optimize oxygen content ($d$) because oxygen content affect crystal structure (orthorhombicity) and superconducting properties of $RE$Ba$_2$Cu$_3$O$_{7-d}$. A high superconducting property is generally achieved in an orthorhombic phase in the system, we estimated the orthorhombicity parameter ($OP$), which is given by $2|a\text{-}b|/(a+b)$, and plotted the estimated $T_c$ and magnetic $J_c$ ($T$ = 2 K, $B$ = 1 T) for $RE$Ba$_2$Cu$_3$O$_{7-d}$ as a function of $OP$ as shown in Fig. 8. Note that the data points are colored according to the number of elements contained at the RE site. From the plot, it is found that Tc does not show a remarkable correlation with $\Delta S_{mix}$ (RE site) and exhibits a clear correlation with $OP$. $J_c$ also exhibits a trend of improvement with increasing $OP$. These facts suggest that disorder introduced by high-entropy alloying at the $RE$ site (see Fig. 4(d)) does not largely affect superconducting properties of $RE$Ba$_2$Cu$_3$O$_{7-d}$, which is a two-dimensional layered compound. Because the trend is clearly different to that observed for cubic (NaCl-type) tellurides with a HEA-type site, crystal-structure dimensionality is a key factor to how the introduction of HEA-type site affects superconducting properties in compounds.

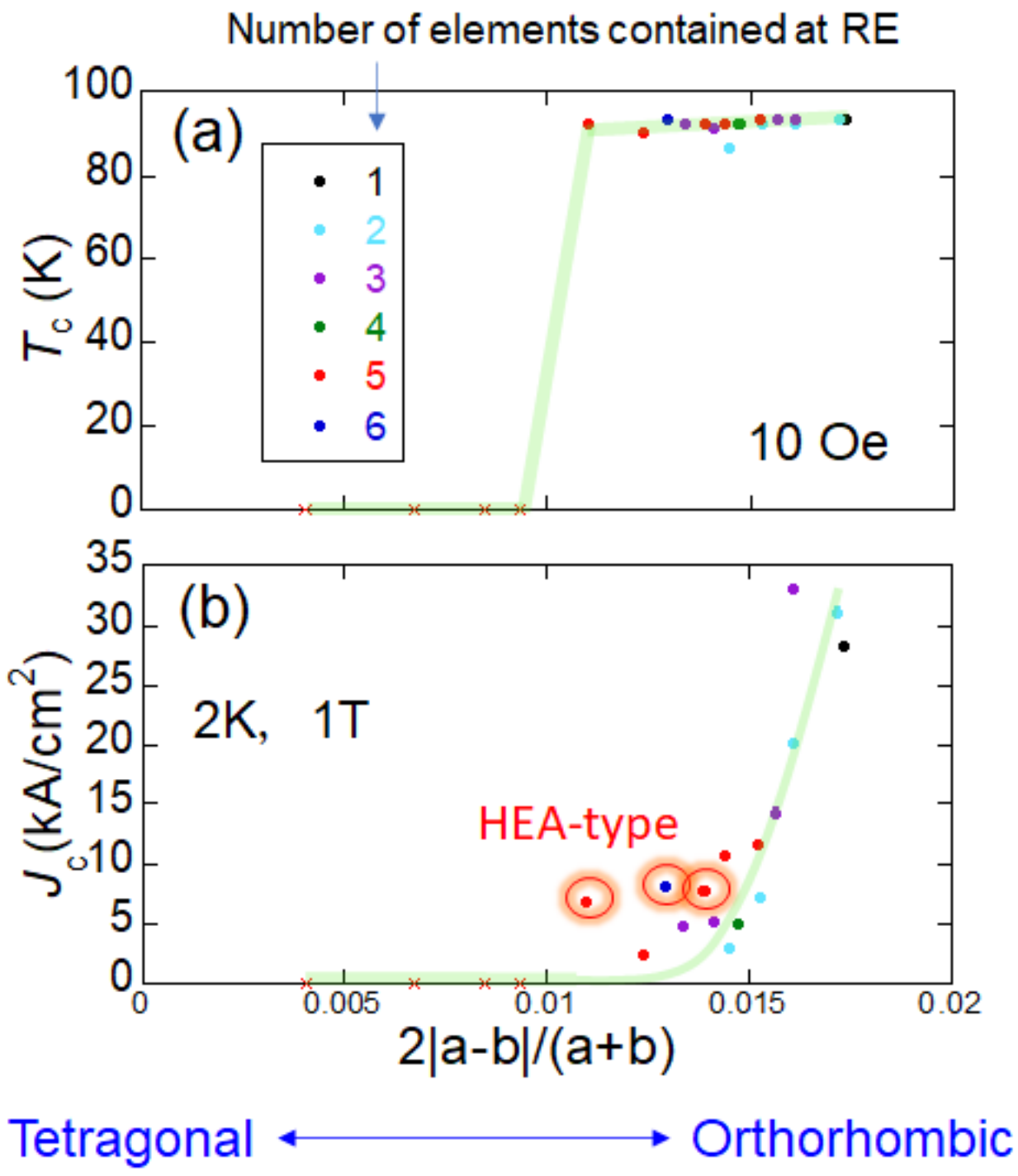

**Fig. 8. Orthorhombicity parameter (2|*a*-*b*|/(*a*+*b*)) dependences of (a) $T_c$ and (b) magnetic $J_c$ ($T$ = 2 K, $B$ = 1 T) for $RE$Ba$_2$Cu$_3$O$_{7-d}$. Original data has been published in Refs. 36. Copyright 2020 by Elsevier.**

In Fig. 8(b), we found three data points for HEA-type samples show a $J_c$ larger than that for the other low-entropy samples at *OP* = 0.01–0.015. Although it has not been fully clarified whether the slightly large $J_c$s in the HEA-type samples are caused by high-entropy alloying or not, the effect of high-entropy alloying for cuprates should be further studied to find the way to improve practical performance of cuprate superconductors.

## 5. BiS$_2$-based layered superconductors *RE*(O,F)BiS$_2$

BiS$_2$-based superconductor family is one of the layered superconductor families and was discover in 2012 [49–51]. The crystal structure is composed of alternate stacks of a conducting BiS$_2$ bilayer and a blocking layer (for example, a *RE*O layer), which is similar to that of high-$T_c$ systems. Furthermore, unconventional superconductivity has been proposed from theoretical and experimental studies on the BiS$_2$-based compounds [51].

A typical BiS$_2$-based system is *RE*OBiS$_2$ (see Fig. 4(e)). Because non-doped REOBiS$_2$ is a semiconductor, electron carrier doping is required to induce metallicity [50]. For *RE* = La, a superconducting transition was observed at 2.5 K after electron doping through partial substitution of O by F in $LaO_{0.5}F_{0.5}BiS_2$. However, the superconductivity states in La(O,F)$BiS_2$ is not bulk in nature. This is due to the presence of the local disorder due to Bi lone pairs, and the local disorder could be suppressed by in-plane chemical pressure effects [52–54]. In-plane chemical pressure can be generated by *RE*-site substitution by smaller *RE* ions or Se substitution for the S site. By increasing in-plane chemical pressure, local disorder is suppressed, and bulk superconductivity is induced [52,53]. Therefore, one can say that *RE*(O,F)$BiS_2$ is a useful system to investigate the effect of structural modification on local crystal structure and superconducting properties. This suggests that the investigation of the effects of introduction of a HEA site in *RE*(O,F)$BiS_2$ would provide us with key information about interlayer interaction through the HEA states.

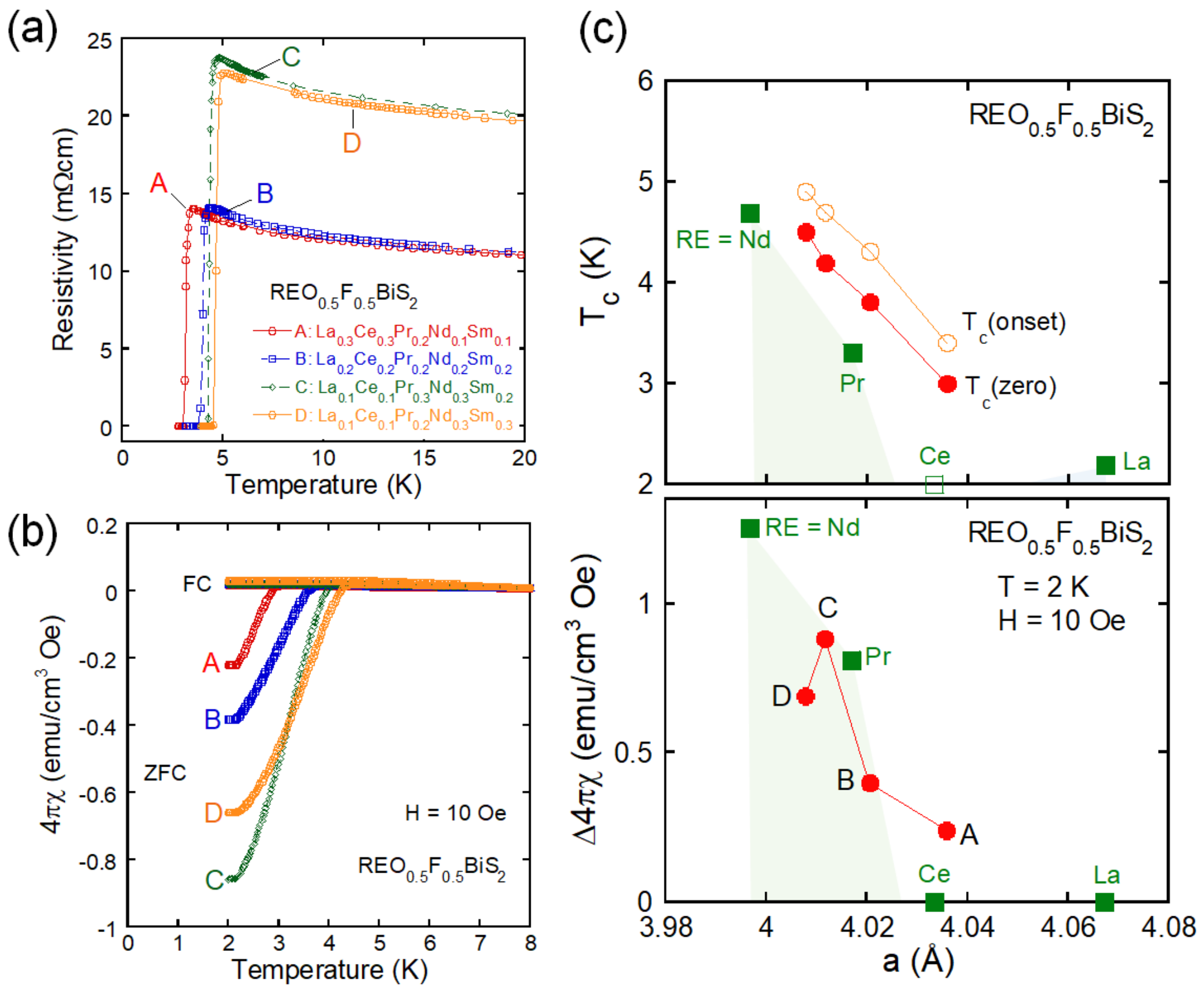

**Fig. 9. (a,b) Temperature dependences of (a) electrical resistivity and (b) magnetic susceptibility for HEA-type $REO_{0.5}F_{0.5}BiS_2$. (c) Lattice constant *a* dependences of $T_c$ and $\Delta 4\pi\chi$ ($T$ = 2 K). Original data has been published in Refs. 37. Copyright 2020 by IOP.**

The HEA-type samples of *RE*$O_{0.5}F_{0.5}BiS_2$ with *RE* = La, Ce, Pr, Nd, Sm were synthesized by solid-state reaction in an evacuated quartz tube [37]. Figures 9(a) and 9(b) show superconducting properties of four different HEA-type *RE*$O_{0.5}F_{0.5}BiS_2$ samples. The $T_c$ varies depending on the *RE*-site composition. As shown in Fig. 9(c), these four samples have different lattice constants *a*, which is corresponding to different in-plane chemical pressures. Therefore, the variation of $T_c$ can be understood with the in-plane chemical pressure scenario. When focusing on the samples with the same lattice constant $a$ and different mixing entropy, we find that the superconducting properties ($T_c$ and $\Delta 4\pi\chi$) of the HEA-type sample are higher than those of low-entropy samples (Figs. 9(c) and 9(d)). The facts indicate that an increase in $\Delta S_{mix}$ for the *RE* site may positively work in improving superconducting properties. Therefore, we systematically prepared $RE$$O_{0.5}F_{0.5}BiS_2$ samples with the same lattice constant $a$ and different $\Delta S_{mix}$ to investigate the interlayer interaction [38].

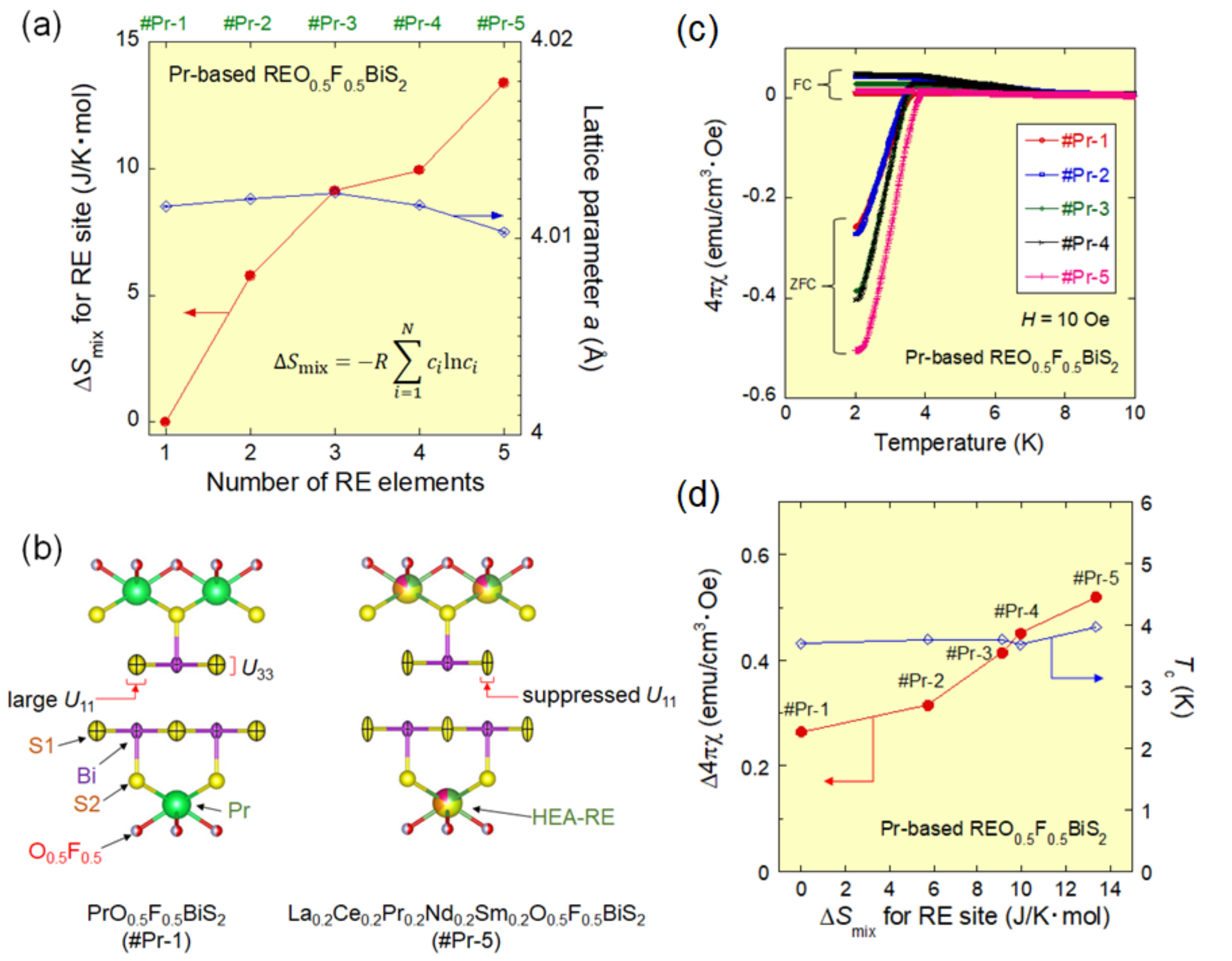


$$\Delta S_{mix} = -R\sum_{i=1}^{N} c_i \ln c_i$$

**Fig. 10. (a) Mixing entropy ($\Delta S_{mix}$) and lattice constant *a* are plotted as a function of the number of RE elements contained in the sample. (b) Schematic images of local structural disorder for low-entropy and HEA-type $REO_{0.5}F_{0.5}BiS_2$. (c) Temperature dependences of $4\pi\chi$. (d) $\Delta S_{mix}$ dependence of $\Delta 4\pi\chi$ ($T$ = 2 K). Original data has been published in Refs. 38. Copyright 2020 by Elsevier.**

As summarized in Fig. 10(a), we succeeded in preparation of a set of five $RE$$O_{0.5}F_{0.5}BiS_2$ samples with almost the same $a$ and systematically varied $\Delta S_{mix}$. Because the lattice constant $a$ nearly corresponds to the degree of the in-plane chemical pressure, the set of samples have almost similar in-plane chemical pressure. Therefore, we could detect the effects of the increase in $\Delta S_{mix}$ to superconducting properties and local structural disorder [38]. Through magnetic susceptibility measurements, we confirmed that shielding volume fraction increases with increasing $\Delta S_{mix}$ while the $T_c$ does not change (Figs. 10(c) and 10(d)). To understand the origin of the improvement of superconducting properties, local structure was analyzed using synchrotron X-ray diffraction. As summarized in Fig. 10(b), suppression of in-plane structural disorder, which was detected by anisotropic displacement parameter for the in-plane direction ($U_{11}$), was achieved by the HEA effect at the *RE* site. Similar trend was observed in another set of $RE\mathrm{O}_{0.5}\mathrm{F}_{0.5}\mathrm{BiS}_2$ samples with larger lattice constant (comparable to that of $\mathrm{CeO}_{0.5}\mathrm{F}_{0.5}\mathrm{BiS}_2$). The results propose that high-entropy alloying at the blocking layer would modify the local structure at the conducting layer in $\mathrm{BiS}_2$-based layered compounds. This effect should be a new strategy to develop novel functional materials with a layered structure.

## 6. Conclusion

In this chapter, we reviewed the works on HEA superconductors and HEA-type superconducting compounds, which have been developed applying the HEA concept in simple alloys to more complicated compounds. By introducing a HEA-type site (or alloying multi sites) in a compound, a high mixing entropy can be achieved. HEA-type compounds would have a high mixing entropy at the alloyed site and a highly (randomly) disordered bonding states. For a three-dimensional system, such as NaCl-type chalcogenides, introduction of the HEA-type *M* site in *MCh* resulted in suppression of $T_c$, but, at the same time, slight enhancements of $H_{c2}$ and $J_c$ were observed. For tetragonal ($\mathrm{CuAl}_2$-type) $Tr\mathrm{Zr}_2$, the introduction of the HEA-type site does not suppress $T_c$. Furthermore, for layered systems, such as $RE\mathrm{Ba}_2\mathrm{Cu}_3\mathrm{O}_{7-d}$ cuprates and $\mathrm{BiS}_2$-based $RE_{0.5}\mathrm{F}_{0.5}\mathrm{BiS}_2$, high-entropy alloying rather improves their superconducting properties. Through investigations of the HEA effects to superconducting and structural properties for HEA-type compounds with different structural types, we conclude that the HEA effects are highly depending on the structural dimensionality. Hence, to effectively use the HEA effect to improve superconducting properties of compounds, target structure should be lower-symmetric, a layered structure or a one-dimensional structure. To obtain further knowledge about the presence/absence of local phase separations, pinning characteristics, and the HEA effect to superconducting pairing states, further investigations on HEA-type superconducting compounds using various probes are required.

## References

1. Onnes H. K. The Superconductivity of Mercury. Comm. Phys. Lab. Univ., Leiden, 1911; 122–124.

2. Adir Moyses Luiz (2011). Room Temperature Superconductivity, Superconductivity - Theory and Applications, Dr. Adir Luiz (Ed.), ISBN: 978-953-307-151-0, InTech, Available from: https://www.intechopen.com/books/superconductivity-theory-and-applications/room-temperature-superconductivity
3. Bednortz J. G., Müller K. A. Possible High $T_c$ Superconductivity in the Ba-La-Cu-O System. Z. Phys. B - Condens. Matter 1986;64 189–193.
4. Takagi H., Uchida S., Kitazawa K., Tanaka S. High-$T_c$ Superconductivity of La-Ba-Cu Oxides. II. –Specification of the Superconducting Phase. Jpn. J. Appl. Phys. 1987;26 L123–L124.
5. Wu M. K., Ashburn J. R., Torng C. J., Hor P. H., Meng R. L., Gao L., Huang Z. J., Wang Y. Q., Chu C. W. Superconductivity at 93 K in a new mixed-phase Y-Ba-Cu-O compound system at ambient pressure. Phys. Rev. Lett. 1987;58 908–910.
6. Schilling A., Cantoni M., Guo J. D., Ott H. R. Superconductivity above 130 K in the Hg-Ba-Ca-Cu-O system. Nature 1993; 363 56–58.
7. Nagamatsu J., Nakagawa N. Muranaka T., Zenitani Y., Akimitsu J. Superconductivity at 39 K in magnesium diboride. Nature 2001;410 63–64.
8. Kamihara Y., Watanabe T., Hirano M., Hosono H. Iron-Based Layered Superconductor $La[O_{1-x}F_x]FeAs$ (x = 0.05−0.12) with Tc = 26 K. J. Am. Chem. Soc. 2008;130 3296–3297.
9. Ren Z. A., Lu W., Yang J., Yi W., Shen X. L., Li Z. C., Che G. C., Dong X. L., Sun L. L., Z. F., Z. Z. X. Superconductivity at 55 K in Iron-Based F-Doped Layered Quaternary Compound $Sm[O_{1-x}F_x]FeAs$. Chin. Phys. Lett. 2008;25 2215–2216.
10. Norman M. R. The Challenge of Unconventional Superconductivity. Science 2011;332 196–200.
11. Drozdov A. P., Eremets M. I., Troyan I. A., Ksenofontov V., Shylin S. I. Conventional superconductivity at 203 kelvin at high pressures in the sulfur hydride system. Nature 2015;525, 73–76.
12. Drozdov A. P., Kong P. P., Minkov V. S., Besedin S. P., Kuzovnikov M. A., Mozaffari S., Balicas L., Balakirev F. F., Graf D. E., Prakapenka V. B., Greenberg E., Knyazev D. A., Tkacz M., Eremets M. I. Superconductivity at 250 K in lanthanum hydride under high pressures. Nature 2019;569 528–531.
13. Somayazulu M., Ahart M., Mishra A. K., Geballe Z. M., Baldini M., Meng Y., Struzhkin V. V., Hemley R. J. Evidence for Superconductivity above 260 K in Lanthanum Superhydride at Megabar Pressures. Phys. Rev. Lett. 2019;122 027001(1–6).
14. Snider E., Dasenbrock-Gammon N., McBride R., Debessai M., Vindana H., Vencatasamy K., Lawler K. V., Salamat A., Dias R. P. Room-temperature superconductivity in a carbonaceous sulfur hydride. Nature 2020;586 373–377.
15. Sun L., Cava R. J. High-entropy alloy superconductors: Status, opportunities, and challenges. Phys. Rev. Mater. 2019;3 090301(1–10).

16. Yeh J. W., Chen S. K., Lin S. J., Gan J. Y., Chin T. S., Shun T. T., Tsau C. H., Chang S. Y. Nanostructured High-Entropy Alloys with Multiple Principal Elements: Novel Alloy Design Concepts and Outcomes. Adv. Eng. Mater. 2004;6 299–303.
17. Tsai M. H., Yeh J. W. High-Entropy Alloys: A Critical Review. Mater. Res. Lett. 2014;2 107–123.
18. Koželj P., Vrtnik S., Jelen A., Jazbec S., Jagličić Z., Maiti S., Feuerbacher M., Steurer W., Dolinšek J. Discovery of a Superconducting High-Entropy Alloy. Phys, Rev. Lett. 2014;113 107001(1–5).
19. Vrtnik S., Koželj P., Meden A., Maiti S., Steurer W., Feuerbacher M., Dolinšek J. Superconductivity in thermally annealed Ta-Nb-Hf-Zr-Ti high-entropy alloys. J. Alloy Compounds 2017;695 3530-3540.
20. Guo J., Wang H., von Rohr F., Wang Z., Cai S., Zhou Y., Yang K., Li A., Jiang S., Wu Q., Cava R. J., Sun L. Robust zero resistance in a superconducting high-entropy alloy at pressures up to 190 GPa. Proc. Natl. Acad. Sci. U.S.A. 2017;114 13144–13147.
21. von Rohr F., Cava R. J. Isoelectronic substitutions and aluminium alloying in the Ta-Nb-Hf-Zr-Ti high-entropy alloy superconductor. Phys. Rev. Mater. 2018;2 034801(1–7).
22. Ishizu N., Kitagawa J. New high-entropy alloy superconductor $Hf_{21}Nb_{25}Ti_{15}V_{15}Zr_{24}$. Results in Phys. 2019;13 102275(1–2).
23. Yuan Y., Wu Y., Luo H., Wang Z., Liang X., Yang Z., Wang H., Liu X., Lu Z. Superconducting $Ti_{15}Zr_{15}Nb_{35}Ta_{35}$ High-Entropy Alloy With Intermediate Electron-Phonon Coupling. Front. Mater. 2018;5 72(1–6).
24. Stolze K., Cevallos F. A., Kong T., Cava R. J. High-entropy alloy superconductors on an α-Mn lattice. J. Mater. Chem. C 2018;6 10441–10449.
25. Wu K. J. Chen S. K., Wu J. M. Superconducting in Equal Molar NbTaTiZr-Based High-Entropy Alloys. Natural Sci. 2018;10 110-124.
26. Nelson W. L., Chemey A. T., Hertz M., Choi E., Graf D. E., Latturner S., Albrecht-Schmitt T. E., Wei K., Baumbach R. E. Superconductivity in a uranium containing high entropy alloy. Sci. Rep. 2020:10 4717(1–8).
27. von Rohr F., Winiarski M. J., Tao J., Klimczuk T., Cava R. J. Effect of electron count and chemical complexity in the Ta-Nb-Hf-Zr-Ti high-entropy alloy superconductor. Proc. Natl. Acad. Sci. U.S.A. 2016;113 7144–7150.
28. Guo J., Lin G., Cai S., Xi C., Zhang C., Sun W., Wang Q., Yang K., Li A., Wu Q., Zhang Y., Xiang T., Cava R. J., Sun L. Record-High Superconductivity in Niobium–Titanium Alloy. Adv. Mater. 2019;31 1807240(1–5).
29. Stolze K., Tao J., von Rohr F., Kong T., Cava R. J. Sc–Zr–Nb–Rh–Pd and Sc–Zr–Nb–Ta–Rh–Pd High-Entropy Alloy Superconductors on a CsCl-Type Lattice. Chem. Mater. 2018;30 906–914.
30. Marik S., Varghese M., Sajilesh K. P., Singh D., Singh R. P. Superconductivity in a new hexagonal

high-entropy alloy. Phys. Rev. Mater. 2019;3 060602(1–6).

31. Mizuguchi Y. Superconductivity in High-Entropy-Alloy Telluride $AgInSnPbBiTe_5$. J. Phys. Soc. Jpn. 2019;88 124708(1–5).
32. Kasem Md. R., Hoshi K., Jha R., Katsuno M., Yamashita A., Goto Y., Matsuda T. D., Aoki Y., Mizuguchi Y. Superconducting properties of high-entropy-alloy tellurides M-Te (M: Ag, In, Cd, Sn, Sb, Pb, Bi) with a NaCl-type structure. Appl. Phys. Express 2020;13 033001(1–4).
33. Yamashita A., Jha R., Goto Y., Matsuda T. D., Aoki Y., Mizuguchi Y. An efficient way of increasing the total entropy of mixing in high-entropy-alloy compounds: a case of NaCl-type $(Ag,In,Pb,Bi)Te_{1-x}Se_x$ (x = 0.0, 0.25, 0.5) superconductors . Dalton Trans. 2020;49 9118–9122.
34. Mizuguchi Y., Kasem Md. R., Matsuda T. D. Superconductivity in $CuAl_2$-type $Co_{0.2}Ni_{0.1}Cu_{0.1}Rh_{0.3}Ir_{0.3}Zr_2$ with a high-entropy-alloy transition metal site. Mater. Res. Lett. 2021;9 141–147.
35. Kasem Md. R., Yamashita A., Goto Y., Matsuda T. D., Mizuguchi Y. Synthesis of high-entropy-alloy-type superconductors $(Fe,Co,Ni,Rh,Ir)Zr_2$ with tunable transition temperature. arXiv:2011.05590.
36. Shukunami Y., Yamashita A., Goto Y., Mizuguchi Y. Synthesis of RE123 high-$T_c$ superconductors with a high-entropy-alloy-type RE site. Physica C 2020;572 1353623(1–5).
37. Sogabe R., Goto Y., Mizuguchi Y. Superconductivity in $REO_{0.5}F_{0.5}BiS_2$ with high-entropy-alloy-type blocking layers. Appl. Phys. Express 2018;11 053102(1–5).
38. Sogabe R., Goto Y., Abe T., Moriyoshi C., Kuroiwa Y., Miura A., Tadanaga K., Mizuguchi Y. Improvement of superconducting properties by high mixing entropy at blocking layers in $BiS_2$-based superconductor $REO_{0.5}F_{0.5}BiS_2$. Solid State Commun. 2019;295 43–49.
39. Fujita Y., Kinami K., Hanada Y., Nagao M., Miura A., Hirai S., Maruyama Y., Watauchi S., Takano Y., Tanaka I. Growth and Characterization of $ROBiS_2$ High-Entropy Superconducting Single Crystals. ACS Omega 2020;5 16819–16825.
40. Heremans J. P., Jovovic V., Toberer E. S., Saramat A., Kurosaki K., Charoenphakdee A., Yamanaka S., Snyder G. J. Enhancement of Thermoelectric Efficiency in PbTe by Distortion of the Electronic Density of States. Science 2008;321 554–557.
41. Tanaka Y., Ren Z., Sato T., Nakayama K., Souma S., Takahashi T., Segawa K., Ando Y. Experimental realization of a topological crystalline insulator in SnTe. Nat. Phys. 2012;8 800–803.
42. Mizuguchi Y., Miura O. High-Pressure Synthesis and Superconductivity of Ag-Doped Topological Crystalline Insulator SnTe ($Sn_{1-x}Ag_xTe$ with x = 0–0.5). J. Phys. Soc. Jpn. 2016;85 053702(1–5).
43. Kobayashi K., Ai Y., Jeschke H. O., Akimitsu J. Enhanced superconducting transition temperatures in the rocksalt-type superconductors $In_{1-x}Sn_xTe$ ($x \leq 0.5$). Phys. Rev. B 2018;97

104511(1–6).
44. Katsuno M., Jha R., Hoshi K., Sogabe R., Goto Y., Mizuguchi Y. High-Pressure Synthesis and Superconducting Properties of NaCl-Type $In_{1-x}Pb_xTe$ (x = 0–0.8). Condens. Matter 2020;5 14(1−10).
45. SuperCon (NIMS database): https://supercon.nims.go.jp/en/
46. Fisk Z., Viswanathan R., Webb G. W. THE RELATION BETWEEN NORMAL STATE PROPERTIES AND $T_c$ FOR SOME $Zr_2X$ COMPOUNDS. Solid State Commun. 1974;15 1797–1799.
47. Teruya A., Kakihana M., Takeuchi T., Aoki D., Honda F., Nakamura A., Haga Y., Matsubayashi K., Uwatoko Y., Harima H., Hedo M., Nakama T., Ōnuki Y. Superconducting and Fermi Surface Properties of Single Crystal $Zr_2Co$. J. Phys. Soc. Jpn. 2016; 85: 034706(1–10).
48. Cai C., Holzapfel B., Hänisch J., Fernández L., Schultz L. High critical current density and its field dependence in mixed rare earth (Nd,Eu,Gd)$Ba_2Cu_3O_{7-\delta}$ thin films. Appl. Phys. Lett. 2004;84 377.
49. Mizuguchi Y., Fujihisa H., Gotoh Y., Suzuki K., Usui H., Kuroki K., Demura S., Takano Y., Izawa H., Miura O. $BiS_2$-based layered superconductor $Bi_4O_4S_3$. Phys. Rev. B 2012;86 220510(1–5).
50. Mizuguchi Y., Demura S., Deguchi K., Takano Y., Fujihisa H., Gotoh Y., Izawa H., Miura O. J. Phys. Soc. Jpn. 2012;81 114725(1–5).
51. Mizuguchi Y. Material Development and Physical Properties of $BiS_2$-Based Layered Compounds. J. Phys. Soc. Jpn. 2019;88 041001 (1–17).
52. Mizuguchi Y., Miura A., Kajitani J., Hiroi T., Miura O., Tadanaga K., Kumada N., Magome E., Moriyoshi C., Kuroiwa Y. In-plane chemical pressure essential for superconductivity in $BiCh_2$-based (Ch: S, Se) layered structure. Sci. Rep. 2015; 5 14968(1–8).
53. Mizuguchi Y., Hoshi K., Goto Y., Miura A., Tadanaga K., Moriyoshi C., Kuroiwa Y. Evolution of Anisotropic Displacement Parameters and Superconductivity with Chemical Pressure in $BiS_2$-Based $REO_{0.5}F_{0.5}BiS_2$ (RE = La, Ce, Pr, and Nd). J. Phys. Soc. Jpn. 2018;87 023704 (1–4).
54. Athauda A., Louca D. Nanoscale Atomic Distortions in the $BiS_2$ Superconductors: Ferrodistortive Sulfur Modes. J. Phys. Soc. Jpn. 2019;88 041004 (1–10).

**Acknowledgments**
The author thanks R. Sogabe, Md. R. Kasem, Y. Shukunami, M. Katsuno, K. Hoshi, R. Jha, Y. Goto, T. D. Matsuda, and O. Miura who have contributed on the works reviewed here. The author would like to thank all the coauthors of the works. The works reviewed here were supported by JSPS-KAKENHI (Nos.: 16H04493, 15H05886, 18KK0076), JST-CREST (Nos.: JPMJCR16Q6, JPMJCR20Q4), and Tokyo Metropolitan Government Advanced Research (No.: H31-1).